\documentclass[pra,twocolumn,showkeywords,longbibliography,10pt]{revtex4-1} 
\usepackage{graphicx}
\usepackage{dcolumn}
\usepackage{color}
\usepackage{times}
\usepackage{bm}
\usepackage{amssymb}
\usepackage{amsmath}
\usepackage{epsfig}
\usepackage{epstopdf}
\usepackage{dsfont}
\usepackage{subfigure}
\usepackage{tikz}
\usepackage[colorlinks, citecolor=blue, linkcolor=blue,urlcolor=blue]{hyperref}
\usepackage[mathscr]{euscript}
\usepackage{orcidlink}
\usepackage{comment}
\usepackage{appendix}
\usepackage{array} 
\usepackage{booktabs} 

\makeatletter
\renewcommand{\maketag@@@}[1]{\hbox{\m@th\normalsize\normalfont#1}}
\makeatother

\begin{document}
	\renewcommand{\thefootnote}{\fnsymbol {footnote}}
	
	\title{{The advantages of extended nonreciprocal quantum batteries}}

	\author{Meng-Long Song~\orcidlink{0009-0002-9705-0241}}
	\affiliation{State Key Laboratory of Opto-Electronic Information Acquisition and Protection Technology, School of Physics, Anhui University, Hefei
		230601,  People's Republic of China}
	
	\author{Zan Cao}
	\affiliation{State Key Laboratory of Opto-Electronic Information Acquisition and Protection Technology, School of Physics, Anhui University, Hefei 230601, People's Republic of China}

    \author{Hai-Tao Dong}
	\affiliation{State Key Laboratory of Opto-Electronic Information Acquisition and Protection Technology, School of Physics, Anhui University, Hefei 230601, People's Republic of China}

\author{Si-Yu Zhang}
	\affiliation{State Key Laboratory of Opto-Electronic Information Acquisition and Protection Technology, School of Physics, Anhui University, Hefei 230601, People's Republic of China}

	\author{Xue-Ke Song} 
	\affiliation{State Key Laboratory of Opto-Electronic Information Acquisition and Protection Technology, School of Physics, Anhui University, Hefei 230601,  People's Republic of China}
	
	\author{Liu Ye}
	\affiliation{State Key Laboratory of Opto-Electronic Information Acquisition and Protection Technology, School of Physics, Anhui University, Hefei 230601,  People's Republic of China}

	\author{Dong Wang~\orcidlink{0000-0002-0545-6205}} \email{dwang@ahu.edu.cn}
	\affiliation{State Key Laboratory of Opto-Electronic Information Acquisition and Protection Technology, School of Physics, Anhui University, Hefei
		230601,  People's Republic of China}


	\date{\today}
	
	\begin{abstract}
   {This study investigates the performance of extended nonreciprocal quantum batteries (QBs), as well as its advantages in energy storage and energy transfer compared to reciprocal charging and the original nonreciprocal batteries. After analyzing the detuning between the charging system and the external pump, we discover that resonance is a key factor in maintaining high-energy batteries and high charging power; furthermore, the detuning of the charger or battery determines the stability of the charging process for different structures. Research on steady-state energy storage in batteries revealed that single-threaded or multi-threaded charging can achieve nearly infinite energy storage in weakly localized environments, thereby demonstrating the significant energy advantages of extended nonreciprocal quantum batteries. Finally, by considering the energy distribution within the charging system, we observe that nonreciprocal charging offers energy transfer advantages unmatched by reciprocal charging; the former achieves a comprehensive balance between charging cost and energy storage capacity that the latter cannot match. As a novel and superior charging protocol, our findings are expected to provide a potent reference for the promotion and practical implementation of nonreciprocal charging.}	
	\end{abstract}
	
	\maketitle
 {Energy transfer and storage are key issues in industry. Current research focuses on the interaction of quantum systems to achieve efficient energy transmission \cite{applied1,applied2,applied3,applied4,applied5}. Quantum batteries (QBs) \cite{QB1,QB2,QB3,QB4,QB5,QB6} were proposed by Alicki \textit{et al.} \cite{propose}, which utilize quantum coherence, entanglement, and other quantumness \cite{qn1,qn2,qn3,qn4,qn5,qn6,qn7,qn8,qn9} to break through traditional limitations. Typical models include the Dicke model \cite{dick1,dick2,dick3,dick4,dick5,dick6,dick7} and spin model \cite{spin1,spin2,spin3,spin4,spin5,spin6}, etc. \cite{other1,other2,other3,other4,other5,other6,other7}, and have been verified on multiple platforms  \cite{plat1,plat2,plat3,plat4}. Given that dissipation significantly affects the energy transmission of open quantum systems \cite{disspation1,disspation2,disspation3,disspation4,disspation5,disspation6,disspation7,disspation8}, it is crucial to isolate the system or control the environmental dissipation to reduce losses. Nonreciprocity violates the Lorentz reciprocity theorem and represents a unique form of energy transfer \cite{nr1,nr2,nr3}. Existing studies have introduced it into quantum battery protocols to guide energy flow, enhance energy storage, and prevent backflow \cite{nrQB1,nrQB2,nrQB3,refer1,refer2,refer3,refer4}. However, research on the multi-charger scheme is still scarce, so this paper further explores its charging advantages. This work expanded the nonreciprocal charging scheme to a single-battery with multi-charger enhancement protocol, and investigate the energy storage performance in single/multi-thread scenarios. The improved scheme significantly enhanced the energy storage capacity, charging power and energy efficiency, and reduced the charging cost. It can provide a theoretical basis for the optimization of actual charging architectures such as optoelectromechanical systems \cite{practice1,practice2}.}

\begin{figure}
		
			\centering
			\subfigure{\includegraphics[width=8.5cm]{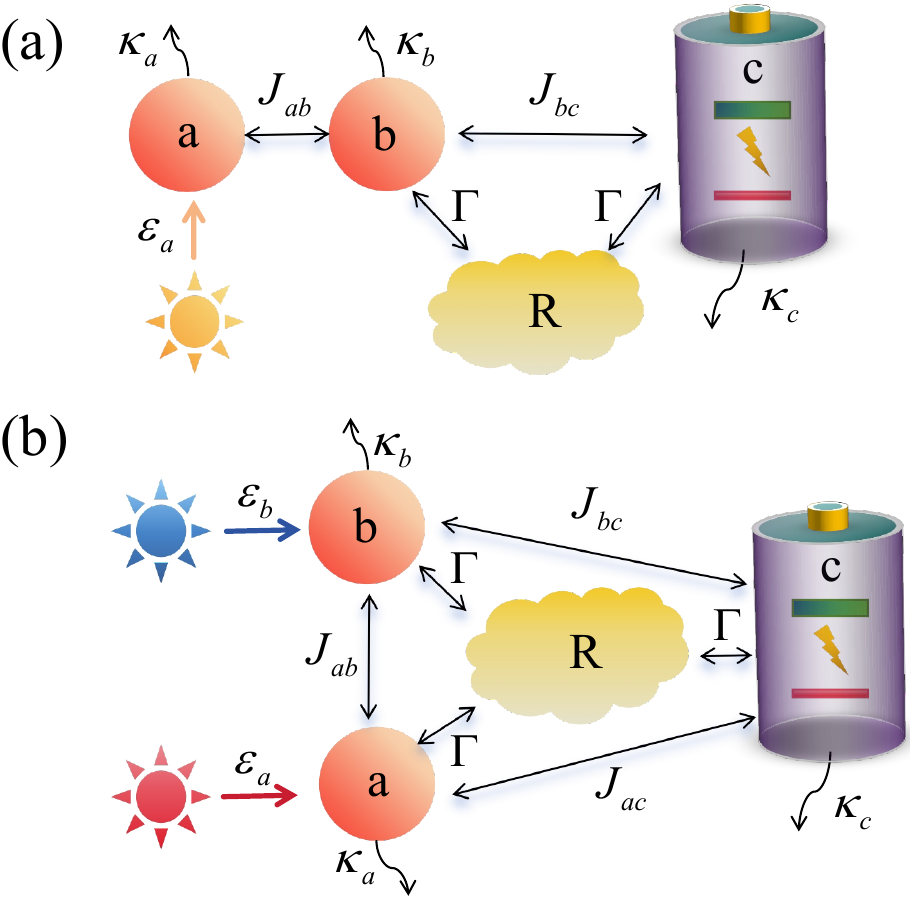}}
		\caption{ {Extended nonreciprocal QB: The subsystems are connected through coherent coupling $J_{ij}$, $ij=\{ab,ac,bc\}$ are connected to the shared reservoir ($R$) at a rate $\Gamma$, and dissipate independently at a rate $\kappa_i$, $i=\{a,b,c\}$; external pumps ($\omega_L$, ${\varepsilon _i}$) drive the charger. The \textit{single-threaded} chain structure only drives the remote charger, and the battery connects to the terminal charger; the \textit{multi-threaded} star structure simultaneously drives all chargers and provides independent power supply.}}
		\label{fig1}
	\end{figure}

 {The extended nonreciprocal system consists of chargers ($a$, $b$)  and battery ($c$). Energy is dissipated into the shared reservoir ($R$) and is affected by the local environment (cf. Fig. \ref{fig1}); for single-thread charging, the external pump drives $a$  through $b$ to charge $c$. The Hamiltonian of the system in the reference frame rotating $\omega_L$ can be reads as ($\hbar  = 1$)}
 \begin{align}
		{H_s} = \sum\limits_{i = \left\{ {a,b,c} \right\}} {{\Delta _i}{i^\dag }i}  + \left( {{J_{ab}}{a^\dag }b + {J_{bc}}{b^\dag }c + {\varepsilon _a}a + \rm{H.c.}} \right).
		\label{Eq.Hs}
	\end{align}
 For multi-threaded protocol, the external pump drives all charger units to charge the battery independently while maintaining connections between units, allowing the Hamiltonian to be expressed as ($\hbar  = 1$)
 \begin{align}
  {H_m} &= \sum\nolimits_{i = \left\{ {a,b,c} \right\}} {{\Delta _i}{i^\dag }i}   \nonumber\\
   &+ \left( {\sum\limits_{ij = \left\{ {ab,bc,ac} \right\}} {{J_{ij}}{i^\dag }j}  + \sum\limits_{i = \left\{ {a,b} \right\}} {{\varepsilon _i}i}  + {\text{H}}.{\text{c}}.} \right),
		\label{Eq.Hm}
	\end{align}
 where ${i = \left\{ {a,b,c} \right\}}$ (${{i^\dag }}$) are the annihilation (creation) operators of the boson modes, satisfying the commutation relation $\left[ {i,{i^\dag }} \right] = 1$. ${\Delta _i} = {\omega _L} - {\omega _i}$ is the detuning between the charging system and the external pump.    

 {Previous studies have shown that adjusting the ratio of $J_{ij}$  and $\Gamma$ can achieve nonreciprocal energy flow \cite{nrQB1,nrQB2}.  For our schemes, nonreciprocal energy injection from the terminal charger or all chargers to the battery can be achieved respectively in the single/multi-thread scenarios.}

Since we are considering the Markovian reservoir, we use the Lindblad quantum master equation to describe the dynamics of the charging system:
\begin{align}
		{\dot \rho _{abc}} =  - i\left[ {H,{\rho _{abc}}} \right] + \sum\limits_{j = a,b,c} {{\kappa _j}{\mathcal{L}_j}\left[ {{\rho _{abc}}} \right]}  + \Gamma {\mathcal{L}_z}\left[ {{\rho _{abc}}} \right],
		\label{Eq.master}
	\end{align}
where ${\mathcal{L}_k}\left[ \rho  \right] = k\rho {k^\dag } - \left\{ {{k^\dag }k,\rho } \right\}/2$. For single-threaded charging, $H = {H_s}$ and $z = \sum\nolimits_{i = b,c} {{p_i}i} $; for multi-threaded charging, $H = {H_m}$ and $z = \sum\nolimits_{i = a,b,c} {{p_i}i} $. Here $p_i$ demonstrates the coupling between the charging system and the shared reservoir. Based on the master equation, we can derive the evolution equation for any operator: $\mathop {\left\langle O \right\rangle }\limits^.  = {\text{Tr}}\left[ {O\dot \rho } \right]$, i.e., 
\begin{align}
		\dot{\langle O\rangle }  =& i\left\langle {\left[ {H,O} \right]} \right\rangle  + {\kappa _j}\left( {\left\langle {{j^\dag }Oj} \right\rangle  - \left\langle {\left\{ {O,{j^\dag }j} \right\}} \right\rangle /2} \right) \nonumber \\
        +& \Gamma \left( {\left\langle {{z^\dag }Oz} \right\rangle  - \left\langle {\left\{ {O,{z^\dag }z} \right\}} \right\rangle /2} \right).
		\label{Eq.operator}
	\end{align}
Taking multi-threaded charging as an example, the evolution equation for the annihilation operator of the charging system is read as
\begin{subequations} 
\begin{align}
   - \mathop {\left\langle a \right\rangle }\limits^.  &= \left( {i{\Delta _a} + \frac{{{\kappa _a} + \Gamma {{\left| {{p_a}} \right|}^2}}}{2}} \right)\left\langle a \right\rangle  + \left( {i{J_{ab}} + \frac{{\Gamma p_a^*{p_b}}}{2}} \right)\left\langle b \right\rangle  \hfill \nonumber \\ 
   &+ \left( {i{J_{ac}} + \frac{{\Gamma p_a^*{p_c}}}{2}} \right)\left\langle c \right\rangle  + i{\varepsilon _a}, \hfill  \label{Eq.o_e1}\\ 
   - \mathop {\left\langle b \right\rangle }\limits^.  &= \left( {iJ_{ab}^* + \frac{{\Gamma {p_a}p_b^*}}{2}} \right)\left\langle a \right\rangle  + \left( {i{\Delta _b} + \frac{{{\kappa _b} + \Gamma {{\left| {{p_b}} \right|}^2}}}{2}} \right)\left\langle b \right\rangle  \hfill  \nonumber\\
   &+ \left( {i{J_{bc}} + \frac{{\Gamma p_b^*{p_c}}}{2}} \right)\left\langle c \right\rangle  + i{\varepsilon _b}, \hfill \label{Eq.o_e2}\\ 
   - \mathop {\left\langle c \right\rangle }\limits^.  &= \left( {iJ_{ac}^* + \frac{{\Gamma {p_a}p_c^*}}{2}} \right)\left\langle a \right\rangle  + \left( {iJ_{bc}^* + \frac{{\Gamma {p_b}p_c^*}}{2}} \right)\left\langle b \right\rangle  \hfill \nonumber\\
   &+ \left( {i{\Delta _c} + \frac{{{\kappa _c} + \Gamma {{\left| {{p_c}} \right|}^2}}}{2}} \right)\left\langle c \right\rangle .  \label{Eq.o_e3}
	\end{align}
    \end{subequations}

\begin{figure}
\begin{minipage}{0.5\textwidth}

		\centering
		\subfigure{\includegraphics[height=3.1cm]{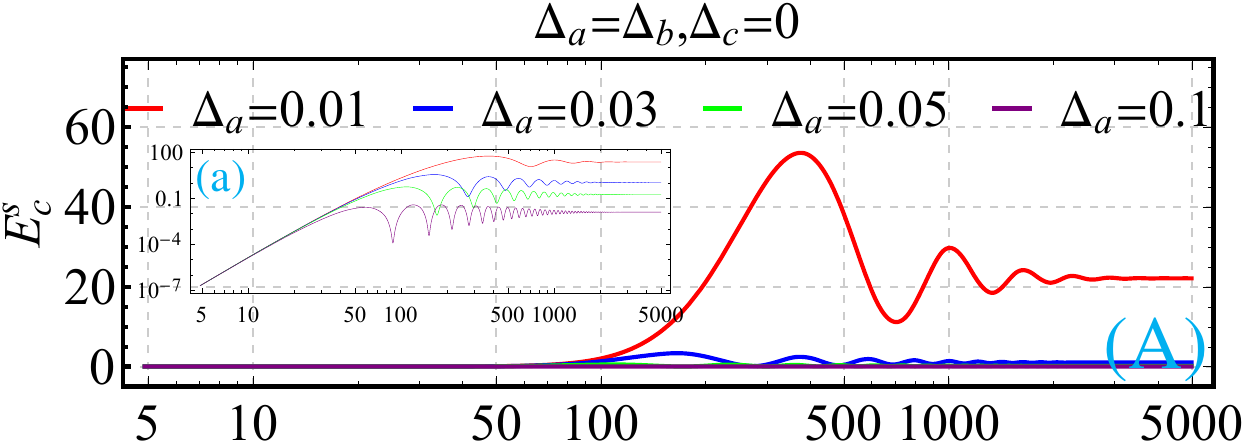}}
		\\
		\subfigure{\includegraphics[height=3cm]{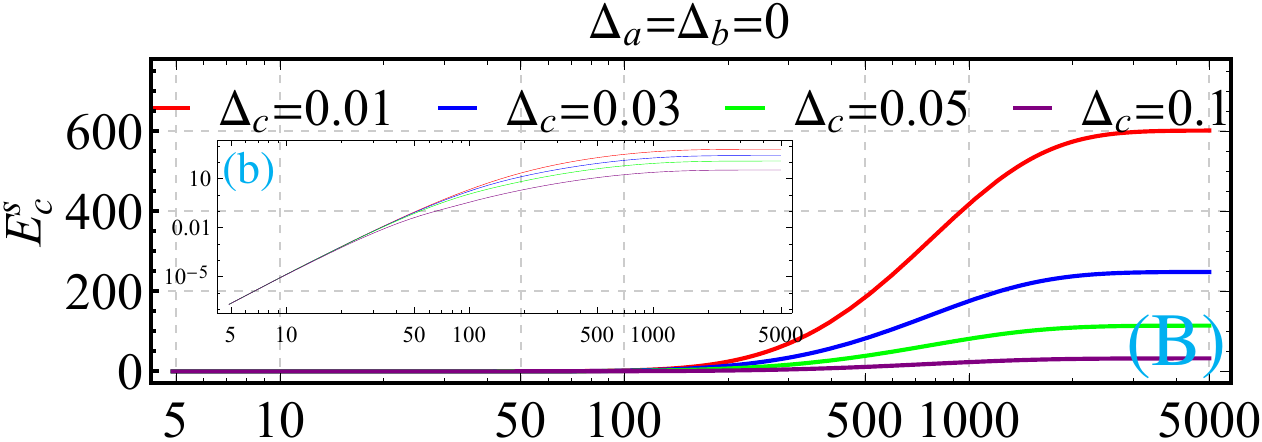}}
		\\
        \subfigure{\includegraphics[height=3cm]{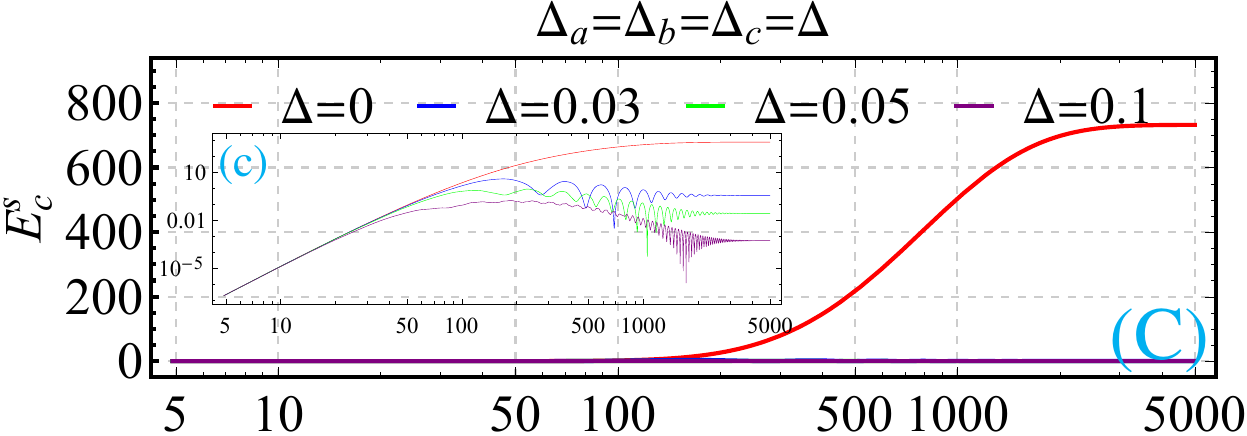}}
		\\
        \subfigure{\includegraphics[height=3.6cm]{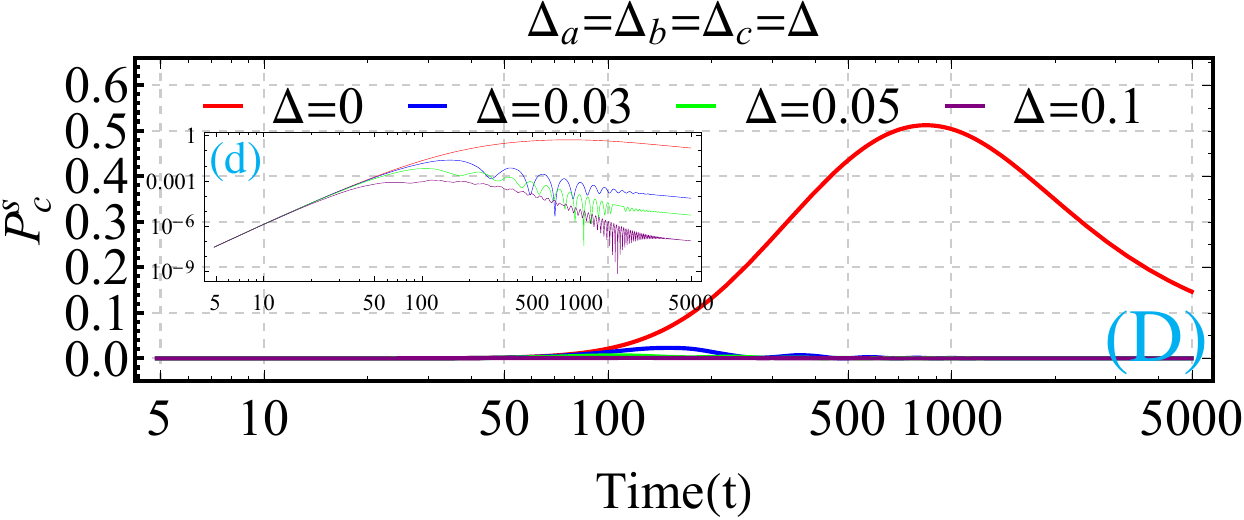}}
		
		\caption{ {For single-threaded scenario: The stored energy $E_c^s$ (unit $\omega_c$)  and charging power $P_c^s=E_c^s/t$ are plotted as a function of time in (A-D, a-d).  Here ${\varepsilon _a} = 0.2$, ${J_{ab}} = \kappa  = 0.003$, $\Gamma  = 0.04,{p_b} = {p_c} = 1$ and ${J_{bc}} = i\Gamma p_b^*{p_c}/2$.}}
		\label{fig2}
        \end{minipage}
    	\end{figure}

 {Simplifying equations Eqs. (\ref {Eq.o_e1})-(\ref {Eq.o_e3}) (setting ${p_a} = {\varepsilon _b} = {J_{ac}} = 0$) leads to the first-order moment of the single-thread charging operator. Under the initial condition of the ground state, the battery energy is denoted as $E_c^{s,m} = {\rm Tr}_{ab}\left[ {{\rho _{abc}}{H_{s,m}}} \right] = {\omega _c}\left\langle {{c^\dag }c} \right\rangle $ (where $ab$ represents the trace of the charger’s degrees of freedom), and the system energy (the second-order moment ${E_j}/{\omega _j} = \left\langle {{j^\dag }j} \right\rangle  = \left\langle {{j^\dag }} \right\rangle \left\langle j \right\rangle  = {\left\langle j \right\rangle ^*}\left\langle j \right\rangle  = {\left| {\left\langle j \right\rangle } \right|^2}$, $\left( {j = a,b,c} \right)$) can be decomposed into the product of the solutions of the first-order moment \cite{disspation1,refer1}. Based on this, the energy distribution, transfer, and steady-state energy (obtained from $\dot{\left\langle O \right\rangle } = 0$) can be calculated.}

Nonreciprocity can now be achieved by balancing coherent coupling and dissipation rates. Concretely, setting ${J_{ac}} = i\Gamma p_a^*{p_c}/2$ and ${J_{bc}} = i\Gamma p_b^*{p_c}/2$ eliminates the $\left\langle c \right\rangle $ term in Eqs. (\ref {Eq.o_e1})-(\ref {Eq.o_e2}),  subsequently enabling multi-threaded nonreciprocal charging.  {Moreover, nonreciprocal charging can be achieved by controlling the coupling between the  $b$ and $c$, ${J_{bc}} = i\Gamma p_b^*{p_c}/2$ for single-threaded charging.}


 {We investigated the impact of charging system detuning on energy transfer and stability in single-thread nonreciprocal protocol: Although charger detuning increases energy robustness (degradation reduction, oscillation slowdown), it leads to a decrease in battery energy storage and a weakening of stability (cf.  Fig. \ref{fig2}(A,a)); battery detuning does not promote energy storage and has almost no effect on stability. At the same level of detuning (e.g., ${\Delta _a} ={\Delta _c}= \left\{ {0.01,0.03,0.05} \right\}$), resonant chargers result in better energy storage than resonant battery (cf. Figs. \ref{fig2}(A)-(B)). Therefore, stable energy storage relies on resonant chargers, and higher energy storage requires both to resonate (cf.  Fig. \ref{fig2}(C,c)). Power  analysis shows that the resonant system has sufficient energy storage and a smooth process, while non-resonant ($\Delta  \ne 0$) systems cause battery nearly failure ($E_c^s,P_c^s \to 0$), and detuned chargers lead to uncontrollable charging and continuous deterioration (cf.  Figs. \ref{fig2}(C)-(D)).}

\begin{figure}
		\begin{minipage}{0.5\textwidth}
			\centering
		\subfigure{\includegraphics[height=3cm]{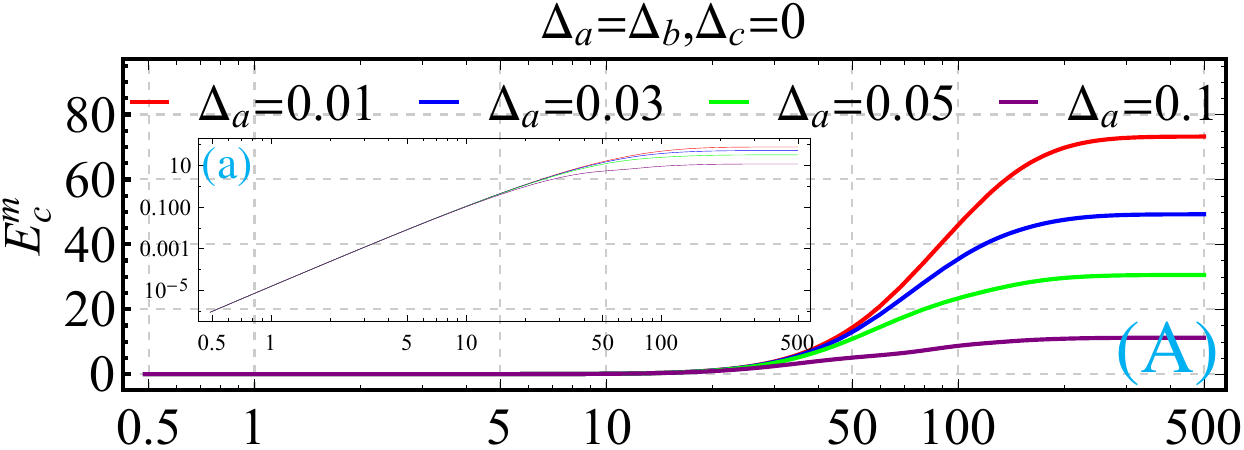}}
		\\
		\subfigure{\includegraphics[height=3cm]{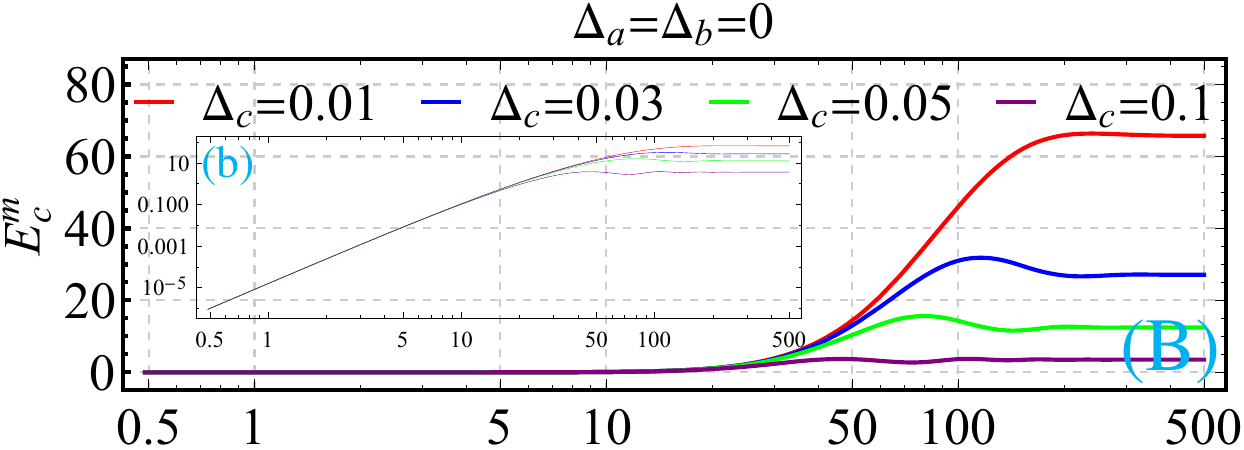}}
        \\
		\subfigure{\includegraphics[height=3cm]{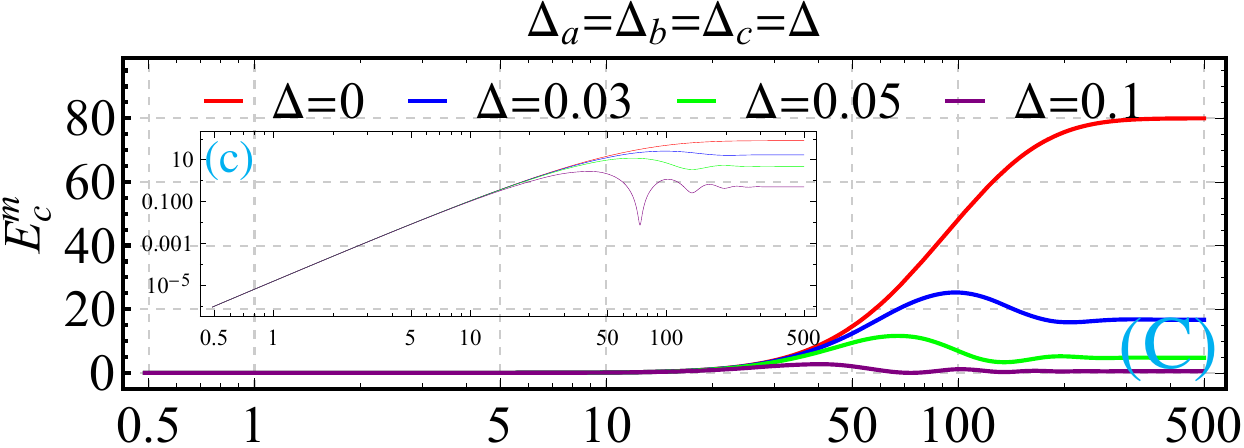}}
        \\
		\subfigure{\includegraphics[height=3.5cm]{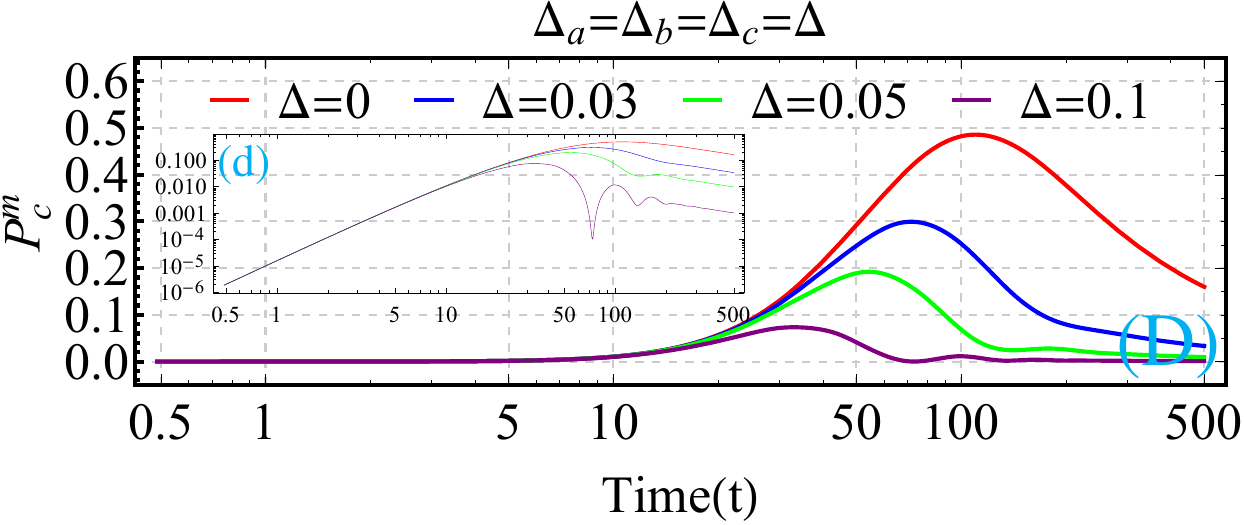}}
		\end{minipage}\hfill
		\caption { {For multi-threaded scenario: The stored energy $E_c^m$ (unit $\omega_c$) and charging power $P_c^m=E_c^m/t$ are plotted as functions of time in (A-D, a-d) Here ${\varepsilon _a} = {\varepsilon _b}= 0.1$, ${J_{ab}} = \kappa  = 0.003$, $\Gamma  = 0.04,{p_a} ={p_b} = {p_c} = 1$,  ${J_{ac}} = i\Gamma p_a^*{p_c}/2$ and ${J_{bc}} = i\Gamma p_b^*{p_c}/2$.}}
		\label{fig3}
	\end{figure}

 {During multi-threaded charging, although the detuning of the charger reduces energy storage, the battery's resistance increases with the increase of detuning (e.g., ${\Delta _a} = \left\{ {0.01,0.03,0.05} \right\}$), and there is no significant fluctuation in charging (cf. Fig. \ref{fig3}(A,a)). Therefore, it is not a critical factor for instability; instead, battery’s detuning not only hinders energy injection but also affects stability (cf. Fig. \ref{fig3}(B,b)), which is different from the single-threaded situation. The resonant system reaches the peak of energy storage and achieves the optimal power (cf. Figs. \ref{fig3}(C)-(D)).}	
	
 {Thereby, the resonance system ensures stable charging and adequate energy storage. Interestingly, multi-thread charging exhibits greater immunity to detuning, both in terms of energy and power (Figs. \ref{fig2}(c)-(d) vs Figs. \ref{fig3}(c)-(d)).}

    \begin{figure}
		\begin{minipage}{0.5\textwidth}
			\centering
		\subfigure{\includegraphics[height=4.3cm]{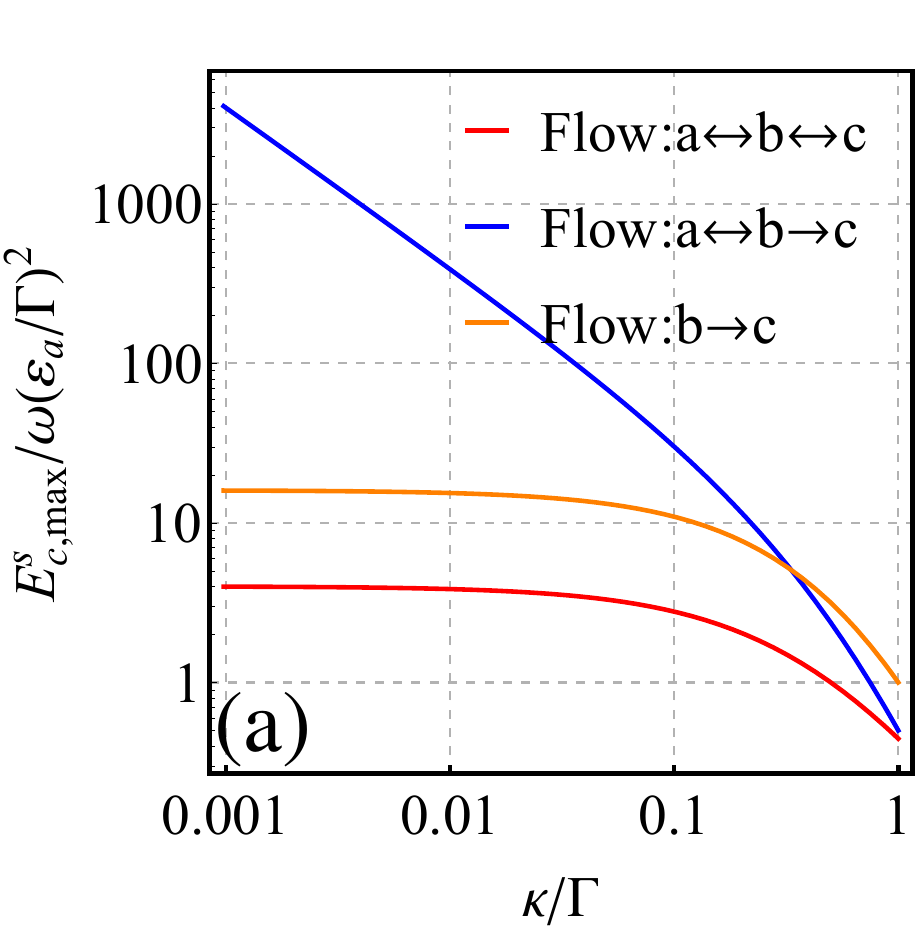}}
		\hspace{0.1cm}
		\subfigure{\includegraphics[height=4.3cm]{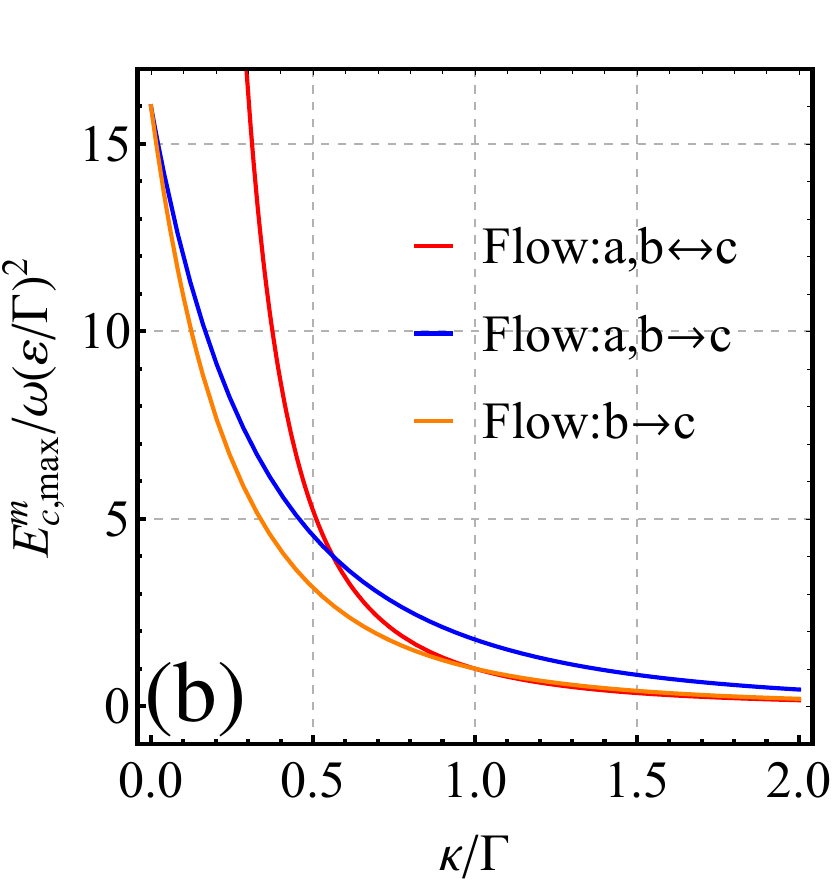}}
		\end{minipage}\hfill
		\caption { {The maximum steady-state stored energy ($E_{c,\max}^s$ and $E_{c,\max}^m$) are plotted as a function of the dissipation ratio $\kappa /\Gamma $. (a) single-threaded charging, (b) multi-threaded charging. The orange line to original nonreciprocal charging \cite{nrQB2}.}}
		\label{fig4}
	\end{figure}

 {Given that nonreciprocal energy storage often reaches its limit in the steady state, we here investigate the steady-state energy storage of the extended protocol and the energy transfer of the system.} For simplicity, we assume that all subsystems (${\omega _a} = {\omega _b} = {\omega _c} = \omega $) have the same local dissipation (${\kappa _a} = {\kappa _b} = {\kappa _c} = \kappa $) and are coupled symmetrically with respect to a shared reservoir (${p_b} = {p_c} = 1$ for single-threaded charging and ${p_a} = {p_b} = {p_c} = 1$ for multi-threaded charging). There is also a same pump (${\varepsilon _a} = {\varepsilon _b} = \varepsilon $, $\Delta  = 0$) supplying power to each charger unit in multi-threaded charging.

 {The existence of nonreciprocity leads to the differences in single-threaded energy storage and flow.  Specifically,  reciprocal  charging  (${J_{bc}} \ne i\Gamma /2$, ${J_{ac}} = {J_{bc}} = J$)  corresponds  to  the free  flow of energy (i.e., $a \leftrightarrow b \leftrightarrow c$), whereas nonreciprocal charging (${J_{bc}} = i\Gamma /2$) results in unidirectional energy injection (i.e., $a \leftrightarrow b \to c$). The different energy flows correspond to different maximum steady-state energies (units of $\omega {\left( {{\varepsilon _a}/\Gamma } \right)^2}$):}
\begin{align}
		E_{c,\max }^s = \left\{ {\begin{array}{*{20}{c}}
  {E_{c,\max }^{s,r} = \frac{4}{{{{(2x + 1)}^2}}},}&{J \to \infty ,} \\ 
  {}&{} \\ 
  {E_{c,\max }^{s,nr} = \frac{4}{{x{{(x + 1)}^3}}},}&{{J_{ab}} = \frac{{\sqrt {\kappa (\Gamma  + \kappa )} }}{2},} 
\end{array}} \right.
		\label{Eq.esc}
	\end{align}
 {where $x = \kappa /\Gamma $.  Eq.  (\ref {Eq.esc})  using the dissipation ratio $\kappa/\Gamma$ as the dependent variable avoids accidental results caused by specific dissipation values. For comparison, the original nonreciprocal charging scheme consists of a charger (mode $b$) and a battery (mode $c$). The nonreciprocity causes the energy to be transmitted in a unidirectional manner ($b \to c$), and its steady-state energy storage can be described as:}
\begin{align}
		E_c^o = \frac{{16}}{{{{(x + 1)}^4}}}.
		\label{Eq.eoc}
	\end{align}
Similarly, we subsequently present reciprocal (${J_{ab}} = {J_{ac}} = {J_{bc}} = J$; $a,b \leftrightarrow c$) and nonreciprocal (${J_{ac}} = {J_{bc}} =i\Gamma /2 $; $a,b \to c$) charging in the multi-threaded framework; these two configurations yield different maximum steady-state energies ${E_{c,\max }^{m,r}}$ and ${E_{c,\max }^{m,nr}}$ (units of $\omega {\left( {{\varepsilon }/\Gamma } \right)^2}$):    
\begin{align}
		E_{c,\max }^m = \left\{ {\begin{array}{*{20}{c}}
  {E_{c,\max }^{m,r} = \frac{{16}}{{{x^2}{{(x + 3)}^2}}},}&{J = 0,} \\ 
  {}&{} \\ 
  {E_{c,\max }^{m,nr} = \frac{{64}}{{{{\left( {{x^2} + 3x + 2} \right)}^2}}},}&{{J_{ab}} = 0.} 
\end{array}} \right.
		\label{Eq.emc}
\end{align}

 {Fig. \ref{fig4}(a) shows a clear advantage of nonreciprocal charging energy storage: the single-thread charging can achieve infinite energy storage (${E_{c,\max }^{s,nr}} \to \infty $) in a weak local environment or under effective isolation ($\kappa  \ll \Gamma $ led to $x  \to 0$); although the original scheme cannot achieve infinite energy storage, its induced capacity is still significantly superior to that of reciprocal charging ${E_{c,\max }^{s,r}}$. Fig.  \ref{fig4}(b) compares multi-thread charging with the original nonreciprocal charging: in a weak local environment ($\kappa  \ll \Gamma $), reciprocal charging can achieve infinite energy storage ($E_{c,\max }^{m,r} \to \infty $); when the coherent effect disappears ($J = 0$), multi-thread reciprocal charging is actually dissipative charging. The results show that when the environment is controllable, any protocol can achieve infinite energy storage. When the environment is restricted, the multi-thread scheme has more advantages and can provide a reference for practical applications.}

\begin{figure}
		\begin{minipage}{0.5\textwidth}
			\centering
		\subfigure{\includegraphics[height=4.3cm]{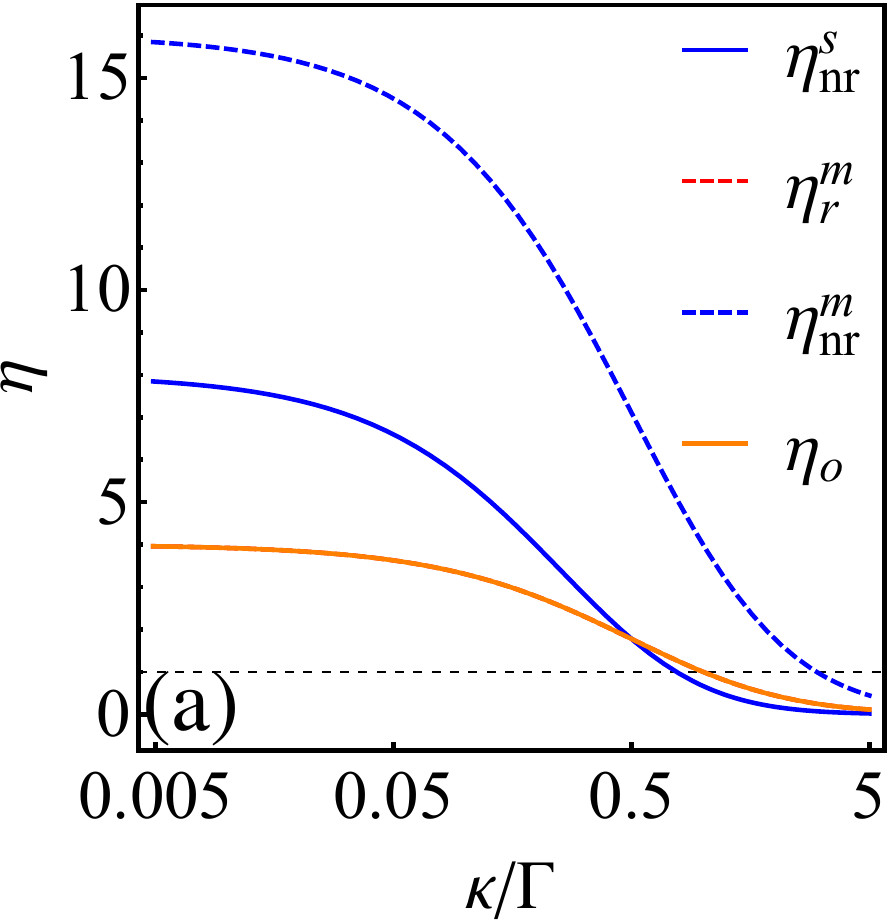}}
		\hspace{0.1cm}
        \subfigure{\includegraphics[height=4.3cm]{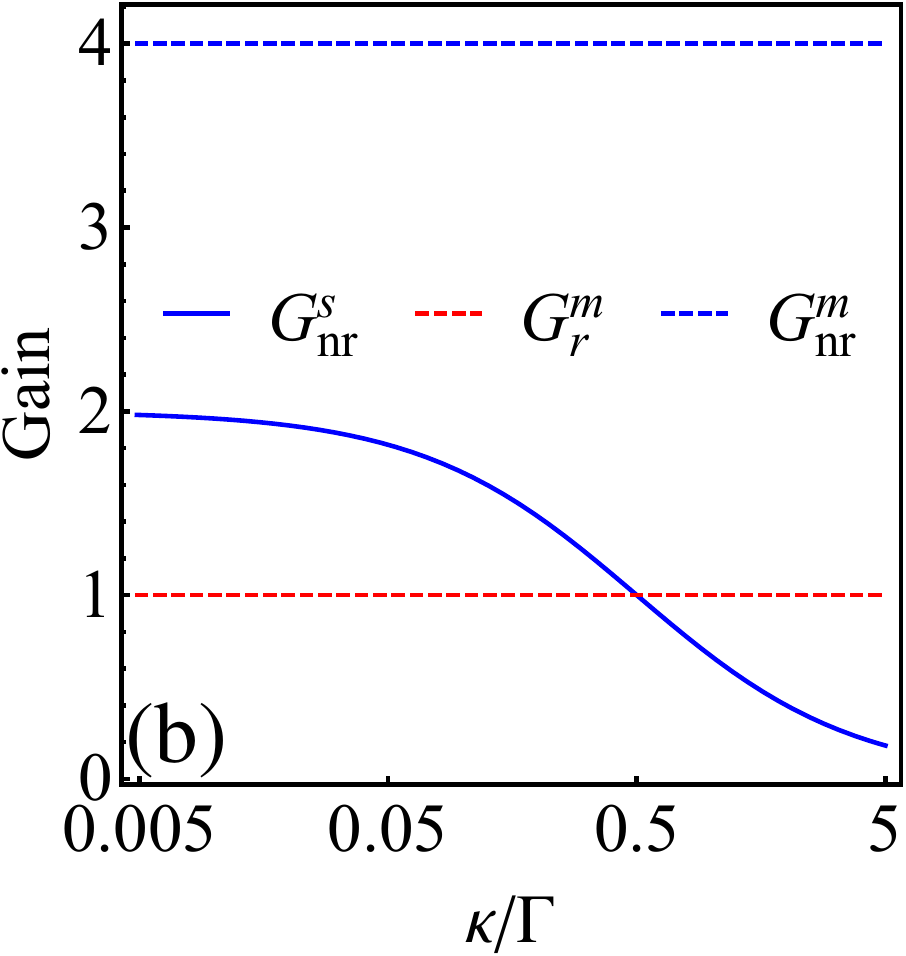}}
		\end{minipage}\hfill
		\caption{ {The energy ratio $\eta $ and gain as functions of the dissipation ratio $\kappa /\Gamma $ are plotted . The black dashed line $\eta=1$ in (a) represents the energy stored equals the charger’s average energy; a higher $\eta $ above it suggest lower-cost charging.}}
		\label{fig5}
	\end{figure}
\begin{table*}
    \centering
\caption{Energy Distribution and Transfer in Charging Systems}
\label{energy}
    \begin{tabular}{cccccc}\hline 
         &  $ {E_c}=E_{c,\max }^{s,r}$ &  ${E_c}=E_{c,\max }^{s,nr}$ & $ {E_c}=E_{c,\max }^{m,r}$ &  $ {E_c}=E_{c,\max }^{m,nr}$ &$ {E_c}=E_c^o$\\\midrule \hline 
 ${E_a}$ &  0 & $1/{x^2}$ & ${\left[ {\left( {2 + 2x} \right)/\left( {3x + {x^2}} \right)} \right]^2}$ & $4/{\left( {x + 2} \right)^2}$ & $-$\\
 ${E_b}$ &  0 & $1/\left( {x + {x^2}} \right)$ & ${\left[ {\left( {2 + 2x} \right)/\left( {3x + {x^2}} \right)} \right]^2}$ & $4/{\left( {x + 2} \right)^2}$ & $4/{\left( {x + 1} \right)^2}$ \\
 $\eta$  &  
 $\eta _r^s \to \infty$ & $\eta _{nr}^s=8/\left[ {{{\left( {x + 1} \right)}^2}\left( {2x + 1} \right)} \right]$ & $\eta_{r}^m= 4/{\left( {x + 1} \right)^2}$ & $\eta_{nr}^m=16/{\left( {x + 1} \right)^2}$ & $\eta_{o}=4/{\left( {x + 1} \right)^2}$\\ \bottomrule
 \hline   \end{tabular}   
\end{table*}

 {The extended nonreciprocal charging requires a comprehensive assessment of the system's energy status and charging costs to avoid increased costs caused by residual energy in chargers. We analyzed the advantages of the expansion scheme and nonreciprocal charging in terms of energy transfer. By defining the energy ratio $\eta  = N{E_c}/\sum\nolimits_{i = a,b} {{E_i}} $ ($N$ is chargers’ number), we eliminated randomness to quantify the transmission effect, and constructed the gain function $G$ to highlight the transmission advantages of the single/multi-thread protocol compared to the original protocol in nonreciprocal and reciprocal scenarios.}

 {After the steady-state storage parameters are determined, the energy of the charging unit can be calculated (cf. Table \ref{energy}). $\eta$ as a dimensionless ratio plotted in Fig. \ref{fig5}(a). The extended nonreciprocal scheme ($\eta_{nr}^s$ and $\eta_{nr}^m$) is higher than the reciprocal and original schemes and has lower cost (higher $\eta$); however, a strong local environment will increase dissipation, causing $\eta<1$, resulting in the failure of low-cost charging.}

 {Although single-threaded reciprocal charging has an $\eta_r^s \to \infty$, it requires extremely strong coherence ($J \to \infty$) and lacks energy storage advantages (cf.  Fig. \ref{fig4}(a)), making the conditions overly stringent. Multi-threaded reciprocal charging, on the other hand, can achieve unlimited energy storage (cf. Fig. \ref{fig4}(b)) but at the cost of a large amount of residual energy in the charger, resulting in extremely high costs. After comprehensive consideration: priority should be given to single-threaded nonreciprocal charging for high energy storage and cost control, while multi-threaded nonreciprocal charging should be selected if cost control is the priority and sufficient energy storage is required.}

 {Fig. \ref{fig5}(b) reveals the energy transfer advantage of nonreciprocal charging, quantified by the energy ratio $G=\eta/\eta_o$ between the extended protocol and the original protocol. The specific effect is as follows:}
\begin{align}
		G = \left\{ {\begin{array}{*{20}{c}}
  {G_r^s \to \infty ,}&{\eta  = \eta _r^s,} \\ 
  {}&{} \\ 
  {G_{nr}^s = \frac{2}{{2x + 1}},}&{\eta  = \eta _{nr}^s,} \\ 
  {}&{} \\ 
  {G_r^m = 1,}&{\eta  = \eta _r^m,} \\ 
  {}&{} \\ 
  {G_{nr}^m = 4,}&{\eta  = \eta _{nr}^m.} 
\end{array}} \right.
		\label{Eq.gain}
	\end{align}
 {Clearly, the extended nonreciprocal protocol's energy transmission advantages is remarkable. Multi-threaded charging can uniquely protect this advantage from environmental interference, and its efficiency always reaches four times that of the original protocol; however, multi-threaded reciprocal charging is not superior to the original protocol in terms of energy transmission due to poor energy distribution.}

 {In summary, this Letter analyzed the energy dynamics of the extended charging protocol and find that the detuning affects the energy storage and power of both single/multi-thread schemes, where the stability of a single-thread charging depends on the detuning of the charger, while that of multi-thread charging depends on the detuning of the battery. When the environment is strictly controlled, both single/multi-thread charging can achieve infinite energy storage, and the extended protocol has a more significant advantage over the original scheme. However, reciprocal charging is difficult to be practical due to strict limitations, insufficient capacity, and high cost; nonreciprocal charging achieves a balance between energy storage capacity and cost: single-thread can achieve controllable cost and infinite energy storage, while multi-thread can ensure sufficient energy storage with the lowest cost.  {Compared to the smooth charging provided by existing multi-auxiliary systems \cite{JPA}, nonreciprocal charging enables low-cost, virtually unlimited energy storage while also achieving stable charging by controlling the resonance between the charging system and the drive, thereby enhancing robustness without increasing system complexity.} This research provides a reference for the promotion of nonreciprocal charging technology.}

\vskip 0.5cm

This work was supported by the National Science Foundation of
China (Grant nos. 12475009, 12075001, and 62471001), Anhui Provincial University Scientific Research Major Project (Grant No. 2024AH040008), the Anhui Provincial Natural Science Foundation (Grant No. 2508085ZD001), and Anhui Province Science and Technology Innovation Project (Grant No.
202423r06050004).	
        



\begin{references}
\bibitem {applied1}  {L. Wang and B. Li, Thermal logic gates: Computation with phonons,} \href{https://doi.org/10.1103/PhysRevLett.99.177208}{Phys. Rev. Lett. {\bf 99}, 177208 (2007).}

\bibitem {applied2}  {A. H. A. Malavazi, B. Ahmadi, P. Mazurek, and A. Mandarino, Detuning effects for heat-current control in quantum thermal devices,} \href{https://doi.org/10.1103/PhysRevE.109.064146}{Phys. Rev. E  {\bf 109}, 064146 (2024).}

\bibitem {applied3}  {S.-Y. Bai and J.-H. An, Floquet engineering to reactivate a dissipative quantum battery,} \href{https://doi.org/10.1103/PhysRevA.102.060201}{Phys. Rev. A  {\bf 102}, 060201 (2020).}

\bibitem {applied4}  {D. Ferraro, M. Campisi, G. M. Andolina, V. Pellegrini, and M. Polini, High-power collective charging of a solid-state quantum battery,} \href{https://doi.org/10.1103/PhysRevLett.120.117702}{Phys. Rev. Lett.   {\bf 120}, 117702 (2018).}

\bibitem {applied5}  {G. M. Andolina, M. Keck, A. Mari, M. Campisi, V. Giovannetti, and M. Polini, Extractable work, the role of correlations, and asymptotic freedom in quantum batteries,} \href{https://doi.org/10.1103/PhysRevLett.122.047702}{Phys. Rev. Lett.   {\bf 122}, 047702 (2019).}

\bibitem {QB1}  { F. Campaioli, F. A. Pollock, F. C. Binder, L. Céleri, J. Goold, S. Vinjanampathy, and K. Modi, Enhancing the Charging Power of Quantum Batteries,} \href{https://doi.org/10.1103/PhysRevLett.118.150601}{Phys. Rev. Lett. {\bf 118}, 150601 (2017).}

\bibitem {QB2}  {F. Campaioli, S. Gherardini, J. Q. Quach, M. Polini, and G. M. Andolina, Colloquium: Quantum batteries,} \href{https://doi.org/10.1103/RevModPhys.96.031001}{Rev. Mod. Phys. {\bf 96}, 031001 (2024).}

\bibitem {QB3}  {J.-Y. Gyhm, D. \ifmmode \check{S}\else \v{S}\fi{}afr\'anek, and D. Rosa, Quantum Charging Advantage Cannot Be Extensive without Global Operations,} \href{https://doi.org/10.1103/PhysRevLett.128.140501}{Phys. Rev. Lett. {\bf 128}, 140501 (2022).}

\bibitem {QB4}  {V. Shaghaghi, V. Singh, G. Benenti and D. Rosa, Micromasers as quantum batteries,} \href{https://iopscience.iop.org/article/10.1088/2058-9565/ac8829}{Quantum Sci. Technol. {\bf 7}, 04LT01 (2022).} 

\bibitem {QB5}  {W.-L. Song, H.-B. Liu, B. Zhou, W.-L. Yang, and J.-H. An, Remote charging and degradation suppression for the quantum battery,} \href{https://doi.org/10.1103/PhysRevLett.132.090401}{Phys. Rev. Lett. {\bf 132}, 090401 (2024).} 

\bibitem {QB6}  {Z.-G. Lu, G.-Q. Tian, X.-Y. Lü, and C. Shang, Topological Quantum Batteries,} \href{https://doi.org/10.1103/PhysRevLett.134.180401}{Phys. Rev. Lett. {\bf 134}, 180401 (2025).}

\bibitem {propose}  {R. Alicki and M. Fannes, Entanglement boost for extractable work from ensembles of quantum batteries,} \href{https://doi.org/10.1103/PhysRevE.87.042123}{Phys. Rev.  E {\bf 87}, 042123 (2013).}

\bibitem {qn1}  {H. L. Shi, S. Ding, Q. K. Wan, X. H. Wang, and W. L. Yang, Entanglement, Coherence, and Extractable Work in Quantum Batteries,} \href{https://doi.org/10.1103/PhysRevLett.129.130602}{Phys. Rev. Lett. {\bf 129}, 130602 (2022).} 

\bibitem {qn2}  {J.-Y. Gyhm and U. R. Fischer, Beneficial and detrimental entanglement for quantum battery charging,} \href{https://doi.org/10.1116/5.0184903}{AVS Quantum Sci. {\bf 6}, 012001 (2024).} 

\bibitem {qn3}  {X. Yang, Y. H. Yang, M. Alimuddin, R. Salvia, S. M. Fei, L. M. Zhao, S. Nimmrichter, and M. X. Luo, Battery Capacity of Energy-Storing Quantum Systems,} \href{https://doi.org/10.1103/PhysRevLett.131.030402}{Phys. Rev. Lett. {\bf 131}, 030402 (2023).} 

\bibitem {qn4}  {G. Francica, F. C. Binder, G. Guarnieri, M. T. Mitchison, J. Goold, and F. Plastina, Quantum Coherence and Ergotropy,} \href{https://doi.org/10.1103/PhysRevLett.125.180603}{Phys. Rev. Lett. {\bf 125}, 180603 (2020).} 

\bibitem {qn5}  {W.-L. Song, J.-L. Wang, B. Zhou, W.-L. Yang, and J.-H. An, Self-discharging mitigated quantum battery,} \href{https://doi.org/10.1103/d9k1-75d4}{Phys. Rev. Lett. {\bf 135}, 020405 (2025).} 

\bibitem {qn6}  {M.-L. Song, L.-J. Li, X.-K. Song, L. Ye, and D. Wang, Environment-mediated entropic uncertainty in charging quantum batteries,} \href{https://doi.org/10.1103/PhysRevE.106.054107}{Phys. Rev. E {\bf 106}, 054107 (2022).} 

\bibitem {qn7}  {M.-L. Song, X.-K. Song, L. Ye, and D. Wang, Evaluating extractable work of quantum batteries via entropic uncertainty relations,} \href{https://doi.org/10.1103/PhysRevE.109.064103}{Phys. Rev. E {\bf 109}, 064103 (2024).} 

\bibitem {qn8}  {M.-L. Song, Z. Cao, X.-K. Song, L. Ye, and D. Wang, Quantum steering as a probe of energy transfer in quantum batteries,} \href{https://doi.org/10.1103/rkyk-14dj}{Phys. Rev. A {\bf 113}, 022209 (2026).} 

\bibitem {qn9}  {A. Sarkar, P. Chaki, P. Ghosh, and U. Sen, Fluctuation in energy extraction from quantum batteries: How open should the system be to control it?,} \href{https://doi.org/10.48550/arXiv.2505.16851}{arXiv:2505.16851 (2025).} 

\bibitem {dick1}  {Y.-Y. Zhang, T.-R. Yang, L. B. Fu, and X. G. Wang, Powerful harmonic charging in a quantum battery
,} \href{https://doi.org/10.1103/PhysRevE.99.052106}{Phys. Rev. E {\bf 99}, 052106 (2019).} 

\bibitem {dick2}  {S. Juli\`{a}-Farr\'{e}, T. Salamon, A. Riera, M. N. Bera, and M. Lewenstein, Bounds on the capacity and power of quantum batteries,} \href{https://doi.org/10.1103/PhysRevResearch.2.023113}{Phys. Rev. Res. {\bf 2}, 023113 (2020).}		

\bibitem {dick3}  {X. Zhang and M. Blaauboer, Enhanced energy transfer in a Dicke quantum battery,} \href{ https://doi.org/10.3389/fphy.2022.1097564}{Front. Phys. {\bf 10}. 1097564 (2023).} 

\bibitem {dick4}  {A. Crescente, M. Carrega, M. Sassetti, and D. Ferraro, Ultrafast charging in a two-photon Dicke quantum battery,} \href{ https://doi.org/10.1103/PhysRevB.102.245407}{Phys. Rev. B {\bf 102}. 245407 (2020).} 

\bibitem {dick5}  {F.-Q. Dou, Y.-Q. Lu, Y.-J. Wang, and J.-A. Sun, Extended Dicke quantum battery with interatomic interactions and driving field,} \href{ https://doi.org/10.1103/PhysRevB.105.115405}{Phys. Rev. B {\bf 105}. 115405 (2022).} 

\bibitem {dick6}  {P. A. Erdman, G. M. Andolina, V. Giovannetti, and F. Noé, Reinforcement Learning Optimization of the Charging of a Dicke Quantum Battery,} \href{ https://doi.org/10.1103/PhysRevLett.133.243602}{Phys. Rev. Lett. {\bf 133}. 243602 (2024).} 

\bibitem {dick7}  {L. Wang, S.-Q. Liu, F.-l. Wu, H. Fan, and S.-Y. Liu, Deep strong charging in a multiphoton anisotropic Dicke quantum battery,} \href{ https://link.aps.org/doi/10.1103/PhysRevA.110.042419}{Phys. Rev. A {\bf 110}. 042419 (2024).} 

\bibitem {spin1}  {F. Q. Dou, H. Zhou, and J. A. Sun, Cavity Heisenberg-spin-chain quantum battery,} \href{https://doi.org/10.1103/PhysRevA.106.032212}{Phys. Rev. A {\bf 106}, 032212 (2022).} 

\bibitem {spin2}  {S. Ghosh and A. Sen(De), Dimensional enhancements in a quantum battery with imperfections,} \href{https://doi.org/10.1103/PhysRevA.105.022628}{Phys. Rev. A {\bf 105}, 022628 (2022).} 

\bibitem {spin3}  {Y. Yao and X. Q. Shao, Optimal charging of open spin-chain quantum batteries via homodyne-based feedback control,} \href{https://doi.org/10.1103/PhysRevE.106.014138}{Phys. Rev. E {\bf 106}, 014138 (2022).} 

\bibitem {spin4}  {W.-X. Guo, F.-M. Yang, and F.-Q. Dou, Analytically solvable many-body Rosen-Zener quantum battery,} \href{https://doi.org/10.1103/PhysRevA.109.032201}{Phys. Rev. A {\bf 109}, 032201 (2024).} 

\bibitem {spin5}  {Z.-B. Niu, Y. Wu, Y.-H. Wang, X. Rong, and J.-F. Du, Experimental Investigation of Coherent Ergotropy in a Single Spin System,} \href{https://doi.org/10.1103/PhysRevLett.133.180401}{Phys. Rev. Lett.  {\bf 133}, 180401 (2024).} 

\bibitem {spin6}  {A. G. Catalano, S. M. Giampaolo, O. Morsch, V. Giovannetti, and F. Franchini, Frustrating Quantum Batteries,} \href{https://doi.org/10.1103/PRXQuantum.5.030319}{PRX Quantum  {\bf 5}, 030319 (2024).} 

\bibitem {other1}  {J. S. Yan and J. Jing, Charging by Quantum Measurement,} \href{https://doi.org/10.1103/PhysRevApplied.19.064069}{Phys.
Rev. Appl. {\bf 19}, 064069 (2023).} 

\bibitem {other2}  {J. Q. Quach and W. J. Munro, Using Dark States to Charge and Stabilize Open Quantum Batteries,} \href{https://doi.org/10.1103/PhysRevApplied.14.024092}{Phys. Rev. Appl. {\bf 14}, 024092 (2020).}

\bibitem {other3}  {F. Barra, K. V. Hovhannisyan, and A. Imparato, Quantum batteries at the verge of a phase transition, } \href{https://iopscience.iop.org/article/10.1088/1367-2630/ac43ed}{New J. Phys.  {\bf 24}, 015003 (2022).}

\bibitem {other4}  {O. Abah, G. De Chiara, M. Paternostro, and R. Puebla, Harnessing nonadiabatic excitations promoted by a quantum critical point: Quantum battery and spin squeezing,} \href{https://doi.org/10.1103/PhysRevResearch.4.L022017}{Phys. Rev. Res.  {\bf 4}, L022017 (2022).}

\bibitem {other5}  {S. Seah, M. Perarnau-Llobet, G. Haack, N. Brunner, and S. Nimmrichter, Quantum Speed-Up in Collisional Battery Charging,} \href{https://doi.org/10.1103/PhysRevLett.127.100601}{Phys. Rev. Lett.  {\bf 127}, 100601 (2021).}

\bibitem {other6}  {D. Rossini, G. M. Andolina, D. Rosa, M. Carrega, and M. Polini, Quantum Advantage in the Charging Process of Sachdev-Ye-Kitaev Batteries,} \href{https://doi.org/10.1103/PhysRevLett.125.236402}{Phys. Rev. Lett. {\bf 125}, 236402 (2020).} 

\bibitem {other7}  {D. Rosa, D. Rossini, G. M. Andolina, M. Polini, and M. Carrega, Ultra-stable charging of fast-scrambling SYK quantum batteries,} \href{ https://doi.org/10.1007/JHEP11(2020)067}{J. High Energy Phys. 11, 067 (2020).} 

\bibitem {plat1}  {J. Q. Quach, K. E. McGhee, L. Ganzer, D. M. Rouse, B. W. Lovett, E. M. Gauger, J. Keeling, G. Cerullo, D. G. Lidzey, and T. Virgili, Superabsorption in an organic microcavity: Toward a quantum battery,} \href{https://www.science.org/doi/10.1126/sciadv.abk3160}{Sci. Adv. {\bf 8}, eabk3160 (2022).} 

\bibitem {plat2}  {I. Maillette de Buy Wenniger, S. E. Thomas, M. Maffei, S. C. Wein, M. Pont, A. Harouri, A. Lema\^{\i}tre, I. Sagnes, N. Somaschi, A. Auff\`eves, and P. Senellart, Experimental Analysis of Energy Transfers between a Quantum Emitter and Light Fields,} \href{https://doi.org/10.1103/PhysRevLett.131.260401}{Phys. Rev. Lett. {\bf 131}, 260401 (2023).} 

\bibitem {plat3}  {D. Ferraro, F. Cavaliere, M. G. Genoni, G. Benenti, and M. Sassetti, Opportunities and challenges 
of quantum batteries,} \href{https://doi.org/10.1038/s42254-025-00906-5}{Nat. Rev. Phys. {\bf 8}, 115 (2026).} 

\bibitem {plat4}  {K. Hymas, J. B. Muir, D. Tibben, J. van Embden, T. Hirai, C. J. Dunn, D. E. Gómez, J. A. Hutchison, T. A. Smith, and J. Q. Quach, Superextensive electrical power from a quantum battery,} \href{https://doi.org/10.1038/s41377-026-02240-6}{Light Sci. Appl.  {\bf 15}, 168 (2026).} 

\bibitem {disspation1}  {R. R. Rodr\'{\i}guez, B. Ahmadi, P. Mazurek, S. Barzanjeh, R. Alicki, and P. Horodecki, Catalysis in charging quantum batteries,} \href{https://doi.org/10.1103/PhysRevA.107.042419}{Phys. Rev. A {\bf 107}, 042419 (2023).} 

\bibitem {disspation2}  {F. Pirmoradian, and K. Mølmer, Aging of a quantum battery,} \href{https://doi.org/10.1103/PhysRevA.100.043833}{Phys. Rev. A {\bf 100}, 043833 (2019).} 

\bibitem {disspation3}  {D. Farina, G. M. Andolina, A. Mari, M. Polini1, and V. Giovannetti, Charger-mediated energy transfer for quantum batteries: An open-system approach,} \href{https://doi.org/10.1103/PhysRevB.99.035421}{Phys. Rev. B {\bf 99}, 035421 (2019).} 

\bibitem {disspation4}  {C. A. Downing and M. S. Ukhtary, Hyperbolic enhancement of a quantum battery,} \href{https://doi.org/10.1103/PhysRevA.109.052206}{Phys. Rev. A {\bf 109}, 052206 (2024).} 

\bibitem {disspation5}  {F. H. Kamin and S. Salimi, Steady-state charging of quantum batteries via dissipative ancillas,} \href{https://doi.org/10.1103/PhysRevA.109.022226}{Phys. Rev. A {\bf 109}, 022226 (2024).} 

\bibitem {disspation6}  {F. T. Tabesh, F. H. Kamin, and S. Salimi, Environment-mediated charging process of quantum batteries,} \href{https://doi.org/10.1103/PhysRevA.102.052223}{Phys. Rev. A {\bf 102}, 052223 (2020).} 

\bibitem {disspation7}  {L. Wang, S.-Q. Liu, F.-l. Wu, H. Fan, N.-N. Li, and Si-Yuan Liu, Global and local performance of a quantum battery under correlated noise channels,} \href{https://doi.org/10.1103/stc3-xkp5}{Phys. Rev. A {\bf 112}, 022206 (2025).} 

\bibitem {disspation8}   {{T. Aziz, M.-L. Song, L. Ye, D. Wang, J. J. Gil, and S. Kais, Spectral bounds on entropy and ergotropy via statistical effective temperature in classical polarization and quantum thermal states,} \href{https://doi.org/10.1103/fhqs-g7c6}{Phys. Rev. Res. {\bf 7}, 033117 (2025).}}

\bibitem {nr1}  {K. M. Case, Transfer problems and the reciprocity principle,} \href{https://doi.org/10.1103/RevModPhys.29.651}{Rev. Mod. Phys. {\bf 29}, 651 (1957).} 

\bibitem {nr2}  {M. Born, Reciprocity Theory of Elementary Particles,} \href{https://doi.org/10.1103/RevModPhys.21.463}{Rev. Mod. Phys. {\bf 21}, 463 (1949).} 

\bibitem {nr3}  {L. Remer, E. Mohler, W. Grill, and B. Lüthi, Nonreciprocity in the optical reflection of magnetoplasmas,} \href{https://doi.org/10.1103/PhysRevB.30.3277}{Phys. Rev. B {\bf 30}, 3277 (1984).} 

\bibitem {nrQB1}  {A. Metelmann and A. A. Clerk, Nonreciprocal Photon Transmission and Amplification via Reservoir Engineering,} \href{https://doi.org/10.1103/PhysRevX.5.021025}{Phys. Rev. X {\bf 5}, 021025 (2015).}

\bibitem {nrQB2}  {B. Ahmadi, P. Mazurek, P. Horodecki, and S. Barzanjeh, Nonreciprocal Quantum Batteries,} \href{https://doi.org/10.1103/PhysRevLett.132.210402}{Phys. Rev. Lett. {\bf 132}, 210402 (2024).} 

 {\bibitem {nrQB3}  {B. Ahmadi, P. Mazurek, S. Barzanjeh, and P. Horodecki, Superoptimal charging of quantum batteries via reservoir engineering: Arbitrary
energy transfer unlocked,} \href{https://doi.org/10.1103/PhysRevApplied.23.024010}{Phys. Rev. Appl. {\bf 23}, 024010 (2025).}}

\bibitem {refer1}  {H.-W. Zhao, Y. Xie , X.-Y. Huang, and G.-F. Zhang, Enhanced charging in multibattery systems by nonreciprocity,} \href{https://doi.org/10.1103/fbv7-m7sd}{Phys. Rev. A {\bf 112}, 022214 (2025).}

\bibitem {refer2}  {C.-Z. Sun, Z.-K. Wang, W.-B. Yan, Y.-J. Zhang, Z.-X. Man, and Q.-Y. Cai, Nonreciprocal charging in a quantum battery via a mediator,} \href{https://doi.org/10.1103/p93y-jflt}{Phys. Rev. A {\bf 112}, 012429 (2025).}

\bibitem {refer3}  {N. A. Khan, X.-Y. Zhang, C.-l. Huang, Y. Liu, and D.-H. He, Collective enhancement in nonreciprocal multimode quantum batteries,} \href{https://doi.org/10.1103/67wh-1fxv}{Phys. Rev. B {\bf 112}, 104318 (2025).}

\bibitem {refer4}  {T. Kalita, M. J. Sarmah, and H. P. Goswami, Thermodynamics of a biophotomimetic nonreciprocal quantum battery,} \href{https://doi.org/10.48550/arXiv.2603.15268}{arXiv:2603.15268 (2026).}

\bibitem {practice1}  {L. D. Toth, N. R. Bernier, A. Nunnenkamp, A. K. Feofanov,
and T. J. Kippenberg, A dissipative quantum reservoir for microwave light using a mechanical oscillator,} \href{https://www.nature.com/articles/nphys4121}{Nat. Phys. {\bf 13}, 787 (2017).}

\bibitem {practice2}  {S. Barzanjeh, M. Wulf, M. Peruzzo, M. Kalaee, P. B.
Dieterle, O. Painter, and J. M. Fink, Mechanical on-chip microwave circulator,} \href{https://www.nature.com/articles/s41467-017-01304-x}{Nat. Commun. {\bf 8}, 953 (2017).}

 {\bibitem {JPA}  {N Behzadi and H Kasani, Mechanism of controlling robust and stable
charging of open quantum batteries,} \href{https://doi.org/10.1088/1751-8121/ac94fc}{J. Phys. A: Math. Theor. {\bf 55}, 425303 (2022).}}

\end{references}
\end{document}